\documentclass[11pt]{article}
 
\input epsf
 
\usepackage{longtable}
\usepackage{amssymb}
\usepackage{mysects}
\usepackage{natbib}
\usepackage{aas_macros}
\usepackage{enumitem}

\setlist{nosep}

\newcommand{\hh}{H$_2$}

\begin{document}

\begin{center}
  {\Large \textbf{Report on the O\,IV and S\,IV lines observed by IRIS}}\\
\medskip
\textbf{\large Peter R.\ Young}\\
  pyoung9@gmu.edu\\
  George Mason University \& NASA Goddard Space Flight
  Center\\
\end{center}

\vspace{1cm}

\noindent\emph{This report has not been submitted to a journal but is provided
  on astro-ph as a reference for other scientists who use IRIS data.}

\bigskip

\noindent The \ion{O}{iv} intercombination lines observed by IRIS between 1397
and 1407~\AA\ provide useful density diagnostics. This document
presents data that address two issues related to these lines:
\begin{enumerate}
\item the contribution of \ion{S}{iv} to the \ion{O}{iv} \lam1404.8
  line; and
\item the range of sensitivity  of the \ion{O}{iv} \lam1399.8/\lam1401.2
  ratio.
\end{enumerate}

\section{Overview}

\ion{O}{iv} has a ground configuration of $2s^22p$, and \ion{S}{iv} a
ground configuration of $3s^23p$ and this similarity leads to common
features in their spectra. In particular, each ion has a set of
intercombination transitions of the form $^2P$--$^4P$ where the upper
state lies in the excited n$s$n$p^2$ configuration. There are five
distinct transitions in these multiplets, and the \ion{O}{iv} lines
lie between 1397 and 1408~\AA, and the \ion{S}{iv} transitions lie
between 1398 and 1424~\AA\ (Table~\ref{tbl.lines}). The \ion{O}{iv} multiplet is significantly
stronger than the \ion{S}{iv} multiplet in most conditions, and both sets
of multiplets show density sensitivity in the $\log\,N_{\rm
  e}$=10--13 range.

\begin{table}[h]
\caption{\ion{O}{iv} and \ion{S}{iv} transitions.}
\begin{center}
\begin{tabular}{cccp{5cm}}
\noalign{\hrule}
\noalign{\smallskip}
\noalign{\hrule}
\noalign{\smallskip}
Ion & Wavelength & Transition & Comment \\
\noalign{\hrule}
\noalign{\smallskip}
\ion{O}{iv} &
   1397.198  & 1/2--3/2 & weak line \\
   &  1399.766  & 1/2--1/2 \\
   &  1401.157  & 3/2--5/2 \\
   &  1404.779  & 3/2--3/2 & blended with \ion{S}{iv} \\
   &  1407.374  & 3/2--1/2 & not observed by IRIS \\
\noalign{\smallskip}
\ion{S}{iv} 
& 1398.040 & 1/2--3/2 & very weak line \\
& 1404.808 & 1/2--1/2 & blended with \ion{O}{iv} \\
& 1406.016 & 3/2--5/2 \\
& 1416.887 & 3/2--3/2 & not observed by IRIS \\
& 1423.839 & 1/2--3/2 & not observed by IRIS  \\
\noalign{\hrule}
\end{tabular}
\end{center}
\label{tbl.lines}
\end{table}

Since \ion{S}{iv} is formed very close in temperature to \ion{O}{iv},
then generally there is little interest in studying the \ion{S}{iv}
lines over the stronger \ion{O}{iv} lines. However, one of the
\ion{S}{iv} lines blends with a useful \ion{O}{iv} transition and so
it is necessary to take account of this (Sect.~\ref{sect.blend}).

The IRIS satellite \citep{iris} observes the
1389.0--1407.0~\AA\ wavelength band, although in practice the upper
limit is $\approx$ 1406.6~\AA. This means that a number of the
transitions are not observed by IRIS (see Table~\ref{tbl.lines} for
details).

There have been many previous studies of the \ion{O}{iv} and
\ion{S}{iv} lines due to their importance in both solar and stellar
spectra, and Table~\ref{tbl.refs} lists some key papers.

\begin{table}[h]
\caption{Previous studies of the \ion{O}{iv} intercombination lines.}
\begin{center}
\begin{tabular}{lp{4in}}
\noalign{\hrule}
\noalign{\smallskip}
\noalign{\hrule}
\noalign{\smallskip}
Paper & Comment\\
\noalign{\hrule}
\noalign{\smallskip}
\citet{cook95} & \ion{O}{iv} and \ion{S}{iv} lines in HRTS solar
spectra and HST/GHRS
spectra of Capella.\\
\citet{harper99} &\ion{O}{iv} and \ion{S}{iv} lines in HST/GHRS
spectra of RR Tel. \\
\citet{teriaca01} & Densities in explosive events from SOHO/SUMER.\\
\citet{keenan02} & \ion{O}{iv} and \ion{S}{iv} lines in
HST/STIS 
spectra of RR Tel, and SOHO/SUMER solar spectra.\\
\citet{keenan09} & \ion{O}{iv} lines in stellar and solar spectra.\\
\noalign{\hrule}
\end{tabular}
\end{center}
\label{tbl.refs}
\end{table}

\section{Atomic data}

Atomic data for \ion{O}{iv} and \ion{S}{iv} used in this work are
obtained from version 8 of the CHIANTI atomic database
\citep{dere97,chianti8}. Radiative decay rates for the
lowest 20 atomic levels (including the levels giving rise to the
intercombination lines) of \ion{O}{iv}  were obtained from \citet{correge04}, and all
other decay rates were taken from \citet{liang-b}. Effective collision
strengths for all transitions were taken from \citet{liang-b},
although an error was corrected in these data, as reported in
\citet{chianti8}.

The \ion{S}{iv} data-set has been unchanged since CHIANTI 5
\citep{chianti5}, and consists of radiative decay rates from
\citet{hibbert02}, \citet{tayal99}, \citet{johnson86} and some
unpublished data of P.R.~Young. The data for the intercombination
transitions are from \citet{hibbert02}. Effective collision strengths
for \ion{S}{iv} are from \citet{tayal00}. 

\section{O\,IV Density sensitivity}

\begin{figure}[h]
\centerline{\epsfxsize=16cm\epsfbox{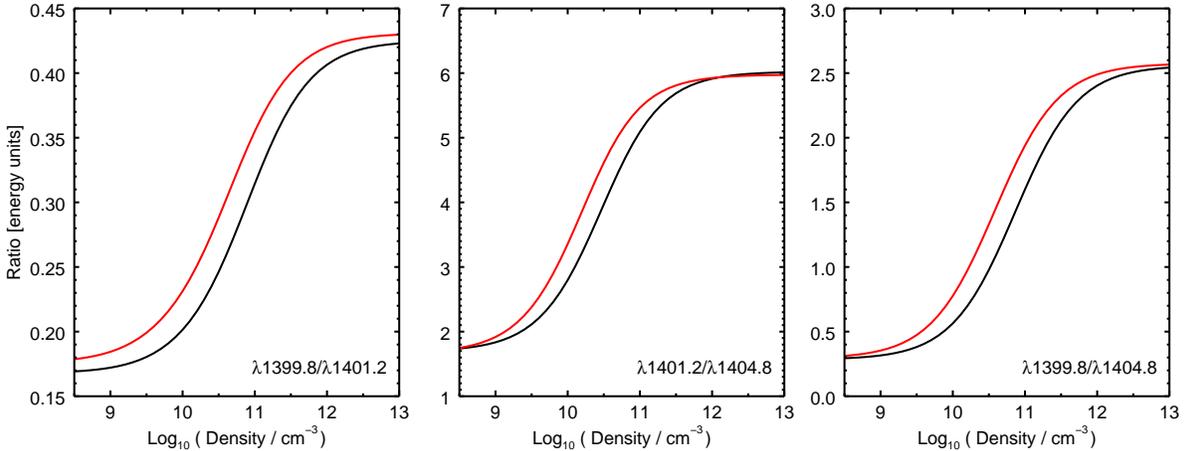}}
\caption{\ion{O}{iv} ratio plots generated with CHIANTI 8. The black
  line has been calculated at $\log\,T=5.15$ and the red line at
  $\log\,T=4.4$. }
\label{fig.ratios}
\end{figure}

Figure~\ref{fig.ratios} shows the density sensitivity of the three
\ion{O}{iv} density diagnostic ratios that are important for
IRIS. Temperature sensitivity is relatively small close to the
temperature of maximum ionization of \ion{O}{iv} ($\log\,T=5.15$), but
in Figure~\ref{fig.ratios} we show the ratios at $\log\,T=4.40$ at
which there are relatively large differences. It is possible in
non-equilibrium conditions that \ion{O}{iv} could be formed closer to
chromospheric temperatures \citep{olluri13}.

\section{IRIS spectra and calibration}

In order to derive densities from the \ion{O}{iv} density diagnostics,
a number of spectra were selected and these are identified in
Table~\ref{tbl.obs}. The spectra were chosen based on criteria such
as: strength of the lines; lack of blending; symmetry of lines; and
possibility of high density. The spectra were created using the IRIS\_SUM\_SPEC routine
in Solarsoft. This routine sums blocks of pixels to create averaged
spectra with error bars. For example, the IDL call to generate
spectrum no.~8 is:
\begin{verbatim}
  IDL> file=iris_find_file('12-oct-2013 21:34')
  IDL> str=iris_sum_spec(file,xpix=9,ypix=[113,115],rnum=1)
\end{verbatim}
The X-pixel and Y-pixel values for each spectra are given in
Table~\ref{tbl.obs}, as well as the raster number (for data-sets in
which multiple rasters were obtained). For each spectrum a background
spectrum was subtracted from the feature spectrum by identifying a
suitable background location near the feature along the slit
direction. The Y-pixel values for the background spectra are given in
Table~\ref{tbl.obs} (the column ``BG Ypix''). Note that the same
X-pixel values as for the feature spectrum were used.

The background spectrum was subtracted from the feature spectrum, with
the error arrays added in quadrature. Emission lines in the
subtracted spectra were
then fit with Gaussians using the IDL routine SPEC\_GAUSS\_WIDGET, which is also available in
Solarsoft.
Line intensities in data number (DN) units derived from the Gaussian fits are given
in Table~\ref{tbl.ints}.

In this work we simply use the line intensities given in DN. The
current version of the IRIS instrument response (implemented through
the IDL routine \verb|iris_get_response|) gives a completely flat
response through the wavelength range of the \ion{O}{iv} lines, and
so the response function has no impact on line ratios. Note
that densities are computed by using theoretical ratios given in
photon units rather than energy units.

\begin{table}[!b]
\caption{Parameters used for computing spectra.}
\begin{center}
  \small
\begin{tabular}{ccccccp{5cm}}
\noalign{\hrule}
\noalign{\smallskip}
\noalign{\hrule}
\noalign{\smallskip}
 & & Rast. & & & BG \\
Index & File start time & no. & Xpix & Ypix & Ypix & Comment \\
\noalign{\hrule}
\noalign{\smallskip}
1 & 11-Oct-2013 23:54 & 0 & 238 & 471:475 & 460:464
     & \ion{S}{iv} strong and fairly symmetric \\
2 & 11-Oct-2013 23:54 & 0 & 76:79 & 714:718 & 733:737 
& Supersonic loop footpoint; fit only supersonic component \\
3 & 11-Oct-2013 23:54 & 0 & 64 & 876:881 & 914:919
     & Bright, elongated structure \\
4 & 11-Oct-2013 23:54 & 0 & 92 & 710:714 & 658:662
& Loop footpoint, not supersonic \\
5 & 10-Sep-2014 11:28 & 0 & 2321 & 338:341 & 344:346
     &Flare kernel site, narrow line \\
6 & 10-Sep-2014 11:28 & 0 & 1988:1992 & 491:492 & 481:485
     &Long loop structure; lines have weak, extended wings \\
7 & 10-Sep-2014 11:28 & 0 & 2381 & 301:304 & 309:313 
     & Flare kernel, narrow lines, extended red wing \\
8 & 24-May-2014 06:10 & 0 & 25:32 & 532:545 & --
     & South coronal hole, just above limb \\
\noalign{\hrule}
\end{tabular}
\end{center}
\label{tbl.obs}
\end{table}

The line intensities derived from the Gaussian fits are given in
Table~\ref{tbl.ints}, and properties derived from the fits are given
in Table~\ref{tbl.prop}. For the \ion{O}{iv} \lam1399.8/\lam1401.2
ratio, we give the measured ratio and the corresponding density,
derived from CHIANTI 8 assuming a temperature of $\log\,T=5.15$. The
measured wavelength separation of the two lines is given in
Table~\ref{tbl.prop} which 
can be compared with the separation given by \citet{young11} of
1.391~\AA. We also give the ratio of the width of \lam1399.8 to that
of \lam1401.2, which gives an indication of the accuracy of the fits
since the two lines should have the same widths. Large differences may
indicate unaccounted for blending.  The final column in
Table~\ref{tbl.prop} is the percentage contribution of \ion{S}{iv} to
the line at 1404.8~\AA\ and this is discussed in the following section.



\section{The 1404.8 line and S\,IV}\label{sect.blend}

The \ion{O}{iv} \lam1404.8 line is potentially a good density
diagnostic relative to \lam1401.1 (Fig.~\ref{fig.ratios}), but there is a known blend with
\ion{S}{iv} \lam1404.8. The \ion{S}{iv} contribution can be estimated
through a branching ratio with \lam1423.8, but this line is not
observed by IRIS. An estimate can be made by considering \lam1406.0,
and this is discussed below.

Based on \emph{Skylab} data, \citet{1979A&A....79..357F} found that \ion{S}{iv}
contributed 10\%\ to the observed feature in quiet Sun conditions, but
this 
rises to 33\%\ in flare conditions. \citet{cook95} state that
\ion{S}{iv} contributes less than 10\%\ by comparing the strength of
the line against the nearby \lam1406.1 transition.
\citet{teriaca01} stated that \ion{S}{iv} only contributed 3.6\%\ to
the observed feature in SUMER spectra but no details were given.

As IRIS does not observe the \ion{S}{iv} \lam1423.8 line, then we only have
the \lam1406.0 line to estimate the \ion{S}{iv} contribution. Using CHIANTI
8 \citep{chianti8}, we find the
\lam1404.8/\lam1406.0 ratio increases slightly with density, with values
of 0.199, 0.202, and 0.224 at $\log\,N_{\rm e}$=9, 10, and 11,
respectively.
In Table~\ref{tbl.prop} we multiply the \lam1406 intensity, where available, by 0.210
and then divide this quantity by the \lam1404.8 intensity to estimate
the percentage contribution of \ion{S}{iv} to the observed feature. We
note that  \citet{cook95} used the same method and they found the same
correction factor (0.21) despite using much older atomic data.

The results demonstrate that \ion{S}{iv} can make a significant
contribution to the blended line and so we generally recommend
\emph{not} to use the 1404.8~\AA\ line as part of a density diagnostic,
unless the \ion{S}{iv} \lam1406.0 is also measured. We note, for
example, that the IRIS flare line list does not include \ion{S}{iv}
\lam1406.0. The \lam1404.8 will be useful in low density data-sets,
such as coronal holes and quiet Sun as \lam1399.8 is weak in these
regions, and so the \lam1401.2/\lam1404.8 ratio should give more
precise results.

\begin{table}[h]
\caption{Line intensities (DN).}
\begin{center}
\begin{tabular}{ccccc}
\noalign{\hrule}
\noalign{\smallskip}
\noalign{\hrule}
\noalign{\smallskip}
& \multicolumn{3}{c}{\ion{O}{iv}} & \ion{S}{iv} \\
\cline{2-4}
\noalign{\smallskip}
Spectrum & \lam1399 & \lam1401 & \lam1404$^b$ & \lam1406 \\
\noalign{\hrule}
\noalign{\smallskip}
1 & $  2576\pm     21$ &$  7499\pm     29$ &$  2239\pm     19$ &$  2738\pm     24$  \\
2 & $   600\pm      8$ &$  2820\pm     11$ &$  1102\pm      9$ &$   179\pm      5$  \\
3 & $   511\pm     13$ &$  1532\pm     15$ &$   622\pm     14$ &$   856\pm     14$  \\
4 & $   256\pm     12$ &$  1093\pm     14$ &$   260\pm     13$ &$    62\pm      8$  \\
5 & $ 10375\pm     42$ &$ 25491\pm     60$ &$  9750\pm     43$ &--- \\
6 & $   184\pm      6$ &$   627\pm      8$ &$   148\pm      6$ &--- \\
7 & $   847\pm     15$ &$  2119\pm     19$ &$   594\pm     15$ &--- \\
8 & $   142\pm      3$ &$   782\pm      3$ &$   360\pm      3$ &$   119\pm      2$  \\

\noalign{\hrule}
\end{tabular}
\end{center}
\label{tbl.ints}
\end{table}

\begin{table}[h]
\caption{Derived properties.}
\begin{center}
\begin{tabular}{cccccc}
\noalign{\hrule}
\noalign{\smallskip}
\noalign{\hrule}
\noalign{\smallskip}
& \multicolumn{4}{c}{\ion{O}{iv} \lam1399.8/\lam1401.2} \\
\cline{2-5}
\noalign{\smallskip}
Spectrum & $\Delta\lambda$ & Intensity ratio & Log Density & Width ratio &\ion{S}{iv} \% \\
\noalign{\hrule}
\noalign{\smallskip}
1 & $  1.379\pm   0.001$ & $  0.343\pm   0.003$ & $ 11.22^{+  0.02}_{-  0.03}$ &$  1.030\pm   0.009$ & $  25.7$   \\
2 & $  1.385\pm   0.001$ & $  0.213\pm   0.003$ & $ 10.16^{+  0.04}_{-  0.03}$ &$  0.993\pm   0.014$ & $   3.4$   \\
3 & $  1.386\pm   0.002$ & $  0.334\pm   0.009$ & $ 11.14^{+  0.07}_{-  0.07}$ &$  0.994\pm   0.027$ & $  28.9$   \\
4 & $  1.388\pm   0.003$ & $  0.234\pm   0.012$ & $ 10.39^{+  0.10}_{-  0.12}$ &$  0.931\pm   0.046$ & $   5.0$   \\
5 & $  1.387\pm   0.000$ & $  0.407\pm   0.002$ & $ 12.02^{+  0.06}_{-  0.04}$ &$  1.022\pm   0.004$ & --- \\
6 & $  1.389\pm   0.002$ & $  0.293\pm   0.011$ & $ 10.85^{+  0.07}_{-  0.07}$ &$  0.897\pm   0.032$ & --- \\
7 & $  1.386\pm   0.001$ & $  0.400\pm   0.008$ & $ 11.86^{+  0.18}_{-  0.13}$ &$  1.030\pm   0.020$ & --- \\
8 & $  1.385\pm   0.001$ & $  0.182\pm   0.003$ & $  9.55^{+  0.10}_{-  0.13}$ &$  0.944\pm   0.017$ & $   7.0$   \\

\noalign{\hrule}
\end{tabular}
\end{center}
\label{tbl.prop}
\end{table}

\section{Blending lines for O\,IV 1399.8}\label{sect.1399}

In this and the following section we consider blending lines for O\,IV
\lam1399.8 and \lam1401.2 that
may affect density measurements from the \lam1401.2/\lam1399.8 ratio.

\begin{figure}[h]
\centerline{\epsfxsize=16cm\epsfbox{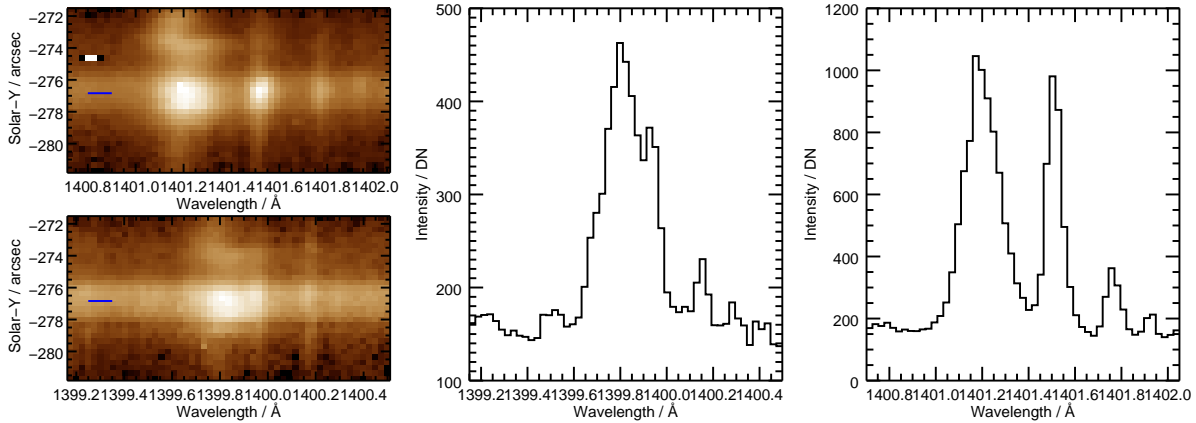}}
\caption{Spectra from an exposure obtained by a raster beginning at 22:08~UT on
  2013 October 12. The left two panels show detector images around the
\ion{O}{iv} \lam1399.8 (upper) and \lam1401.2 (lower) lines, and the
middle and right panel show spectra obtained at Y-pixel 114 (indicated
by the short blue lines on the images.}
\label{fig.o4-fe2}
\end{figure}

\lam1399.8 is partly blended with a \ion{Fe}{ii} line at
1399.960~\AA\ and Fig.~\ref{fig.o4-fe2} shows an example where this
line is quite strong, adding a narrow peak to the long wavelength side
of the \ion{O}{iv} line. Visually the line is about the same strength
as \ion{Fe}{ii} \lam1401.777 which is close to \ion{O}{iv} \lam1401.2 (see
Figure~\ref{fig.o4-fe2}), and so this unblended line may be used to
estimate whether the \ion{Fe}{ii} contribution is important to
\ion{O}{iv}. 

\begin{figure}[h]
\centerline{\epsfxsize=16cm\epsfbox{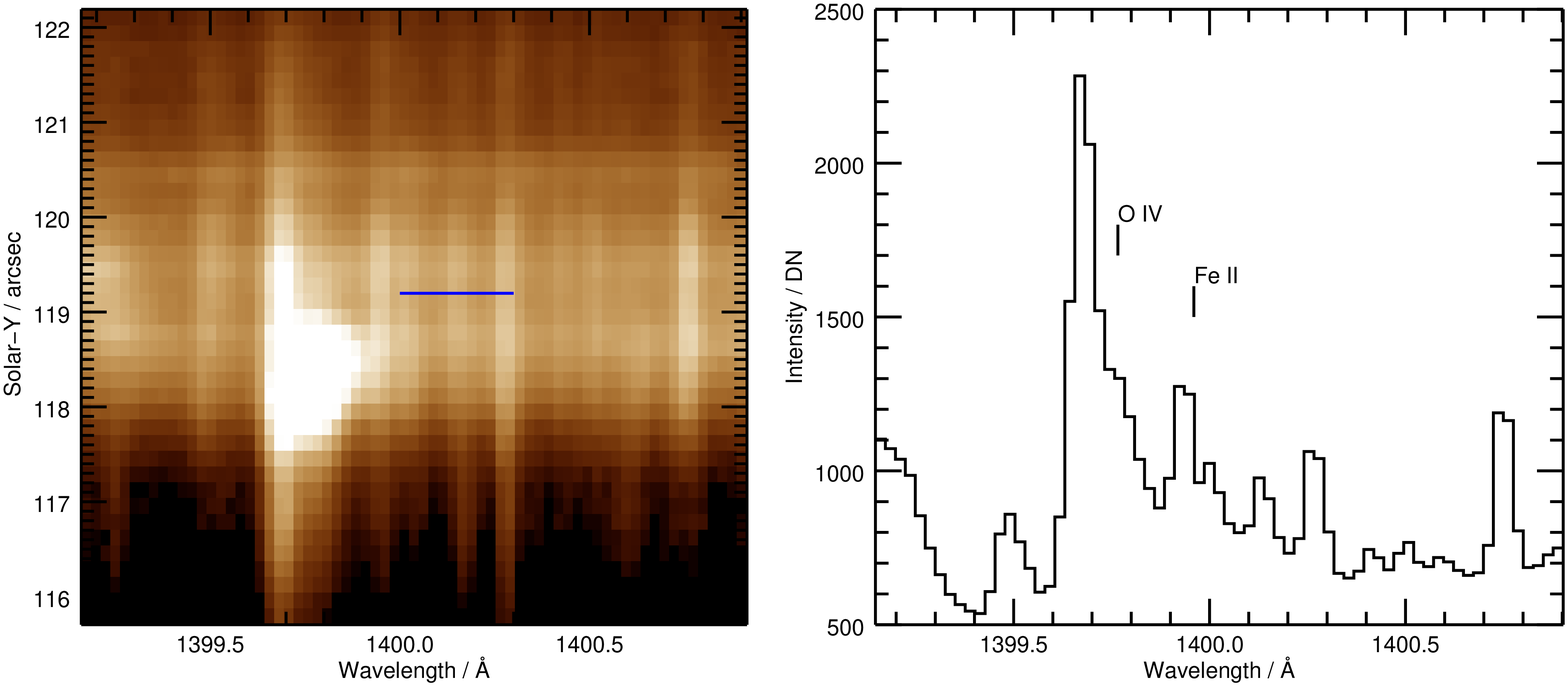}}
\caption{Spectrum for the 2014 September 10 X-flare. An exposure from
  X-pixel 2314 is shown in the left panel, and has been saturated to
  better show the weak lines. The \ion{O}{iv} \lam1399 line is the
  broad, bright structure. The blue line indicates the 1D spectrum
  shown in the right panel, which corresponds to Y-pixel 353.}
\label{fig.1399-blend}
\end{figure}

A more complex blending scenario can occur in flare ribbons, and 
Figure~\ref{fig.1399-blend} shows an example spectrum for the 2014
September 10 X-flare in the vicinity of the \ion{O}{iv} \lam1399
line. The bright blob in the image is the \ion{O}{iv} line, and one
can see a number of very narrow features in the spectrum that have a
larger extent in the Y-direction than the \ion{O}{iv} line. The
strongest of the displayed lines is at 1399.69~\AA\ and, based on the
intensity distribution along the slit, appears to be a \hh\ line but
the identification is not known. Care must be taken when deriving the
intensity of \ion{O}{iv} \lam1399.8 from flare ribbons to estimate the
strength of both this line and \ion{Fe}{ii} \lam1399.96.

\section{Blending lines for O\,IV 1401.1}\label{sect.1401}

A commonly-seen line close to \lam1401.1 is \ion{S}{i} \lam1401.514,
although it is well-separated from the \ion{O}{iv} line and generally
not a problem (Fig.~\ref{fig.o4-fe2}). Note that the velocity of the \ion{S}{i} line relative
to the \ion{O}{iv} line is 76.4~\kms. Near sunspots strong supersonic
downflows of $\approx$~90~\kms\ are often seen in the \ion{O}{iv} line
\citep{straus15}, and so the \ion{S}{i} line can potentially be a
problem when studying such flows, however the author's experience is
that \ion{S}{i} is generally negligible in these structures.

\section{The high density limit}

\begin{figure}[h]
\centerline{\epsfxsize=10cm\epsfbox{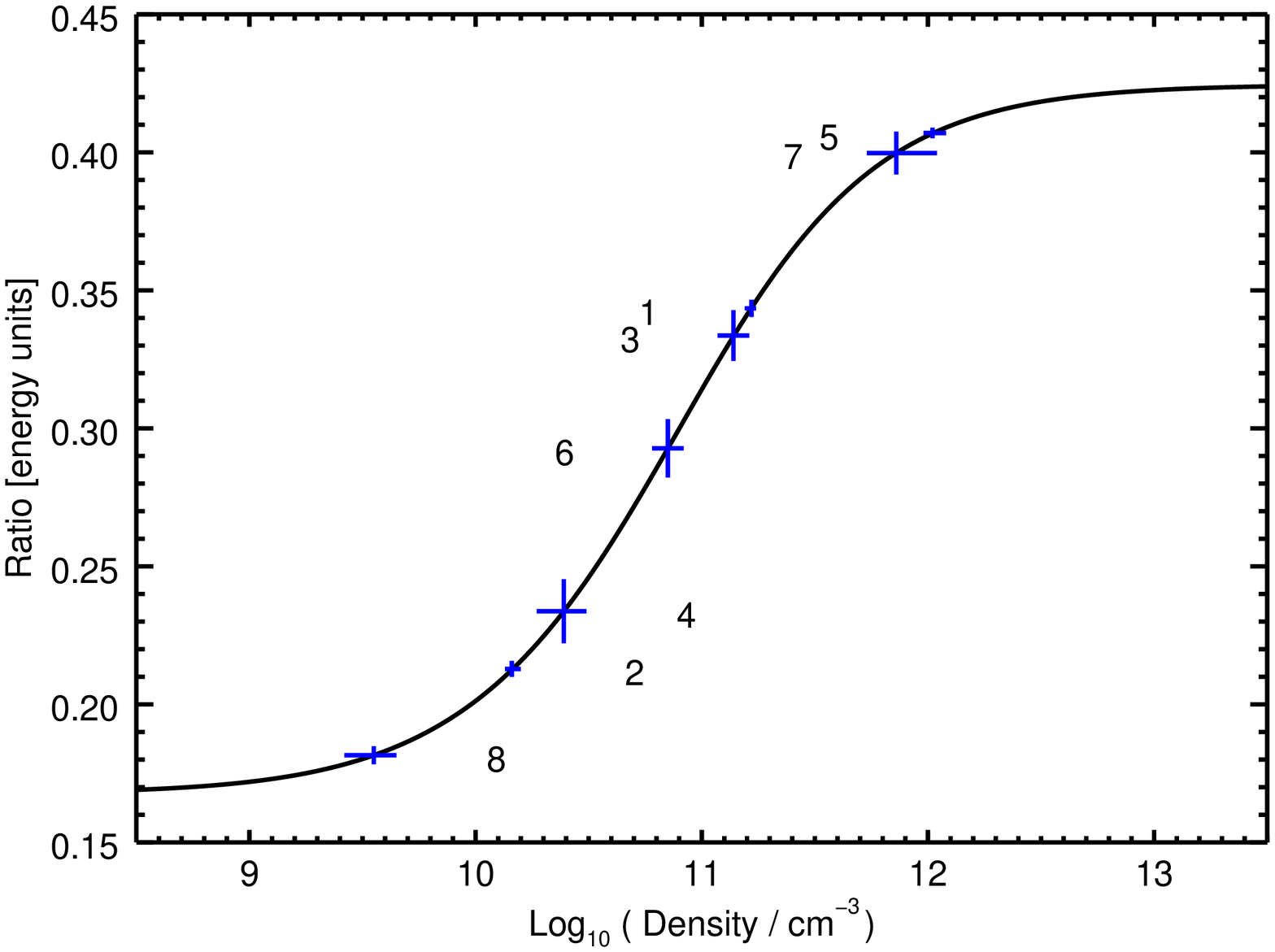}}
\caption{The solid line shows the theoretical variation of the
  \ion{O}{iv} \lam1399.8/\lam1401.2 ratio as a function of density
  computed at $\log\,T=5.15$ using CHIANTI 8. The blue crosses show
  the measured ratios and densities with 1-$\sigma$ error
  bars. Numbers indicate the spectrum from which the measurements were
made (Table~\ref{tbl.obs}).}
\label{fig.dens}
\end{figure}

Figure~\ref{fig.dens} shows the densities derived from  the
\ion{O}{iv} \lam1399.8/\lam1401.2 ratio for each of 
the spectra identified in Table~\ref{tbl.obs}. The data span most of
the density range of the ratio, from the coronal hole observation
(spectrum 8) to the flare kernels (spectra 5 and 7). The fact that
the measured ratios lie within the expected range of variation gives
confidence in the atomic data for \ion{O}{iv}. The 2014 September 10
X-flare does show examples of densities very close to
$10^{12}$~cm$^{-3}$. We note that in features referred to as ``bombs''
by \citet{peter14}, and also in flare kernels the \ion{O}{iv} lines
often become very weak or disappear. This was noted from \emph{Skylab}
data by \citet{feldman78} and was suggested to be due to densities
that are so high that the \ion{O}{iv} emitting levels are de-populated
by electron collisions, reducing the strength of the lines. This would
happen at $\approx 10^{13}$~cm$^{-3}$. The fact that densities of
$\approx 10^{12}$ can be measured with the \ion{O}{iv} ratio in later
stages of flare kernel evolution suggests
that higher densities may be feasible in the early stages of flares.

As a sanity check on the extreme ratio values that were found in this analysis,
Appendix~\ref{sect.sanity} shows over-plots of the two \ion{O}{iv} on top
of each other.

\section{Summary}

The \ion{O}{iv} density diagnostics observed by IRIS have been
investigated and the following results found.
\begin{enumerate}
\item The line observed at 1404.8~\AA\ is dominated by \ion{O}{iv} but
  can contain a contribution from \ion{S}{iv} up to 29\%\ and so it is
  recommended that the \ion{S}{iv} \lam1406.0 line is observed in
  order to estimate the contribution.
  \item The \ion{O}{iv} \lam1399.8 line is blended with an unknown
    line at 1399.69~\AA, which is likely due to \hh. This line can be
    stronger than the \ion{O}{iv} line in flare kernels.
  \item A low density of $\log\,N_{\rm e}=9.5$ is measured just above
    the limb in a coronal hole.
    \item In two flare kernel sites of the 2014 September 10 X-flare
      the density reached $\log\,N_{\rm e}=12$.
\end{enumerate}

\bibliographystyle{apj}
\bibliography{myrefs}

\begin{thebibliography}{}

\bibitem[\protect\citeauthoryear{{Cook} et~al.}{{Cook} et~al.}{1995}]{cook95}
{Cook}, J.~W., {Keenan}, F.~P., {Dufton}, P.~L., {Kingston}, A.~E., {Pradhan},
  A.~K., {Zhang}, H.~L., {Doyle}, J.~G.,  \& {Hayes}, M.~A. 1995, \apj, 444,
  936

\bibitem[\protect\citeauthoryear{{Corr{\'e}g{\'e}} \&
  {Hibbert}}{{Corr{\'e}g{\'e}} \& {Hibbert}}{2004}]{correge04}
{Corr{\'e}g{\'e}}, G.,  \& {Hibbert}, A. 2004, Atomic Data and Nuclear Data
  Tables, 86, 19

\bibitem[\protect\citeauthoryear{{De Pontieu} et~al.}{{De Pontieu}
  et~al.}{2014}]{iris}
{De Pontieu}, B., et~al. 2014, \solphys, 289, 2733

\bibitem[\protect\citeauthoryear{{Del Zanna} et~al.}{{Del Zanna}
  et~al.}{2015}]{chianti8}
{Del Zanna}, G., {Dere}, K.~P., {Young}, P.~R., {Landi}, E.,  \& {Mason}, H.~E.
  2015, ArXiv e-prints

\bibitem[\protect\citeauthoryear{{Dere} et~al.}{{Dere} et~al.}{1997}]{dere97}
{Dere}, K.~P., {Landi}, E., {Mason}, H.~E., {Monsignori Fossi}, B.~C.,  \&
  {Young}, P.~R. 1997, \aaps, 125, 149

\bibitem[\protect\citeauthoryear{{Feldman} \& {Doschek}}{{Feldman} \&
  {Doschek}}{1978}]{feldman78}
{Feldman}, U.,  \& {Doschek}, G.~A. 1978, \aap, 65, 215

\bibitem[\protect\citeauthoryear{{Feldman} \& {Doschek}}{{Feldman} \&
  {Doschek}}{1979}]{1979A&A....79..357F}
{Feldman}, U.,  \& {Doschek}, G.~A. 1979, \aap, 79, 357

\bibitem[\protect\citeauthoryear{{Harper} et~al.}{{Harper}
  et~al.}{1999}]{harper99}
{Harper}, G.~M., {Jordan}, C., {Judge}, P.~G., {Robinson}, R.~D., {Carpenter},
  K.~G.,  \& {Brage}, T. 1999, \mnras, 303, L41

\bibitem[\protect\citeauthoryear{{Hibbert}, {Brage}, \& {Fleming}}{{Hibbert}
  et~al.}{2002}]{hibbert02}
{Hibbert}, A., {Brage}, T.,  \& {Fleming}, J. 2002, \mnras, 333, 885

\bibitem[\protect\citeauthoryear{{Johnson}, {Kingston}, \& {Dufton}}{{Johnson}
  et~al.}{1986}]{johnson86}
{Johnson}, C.~T., {Kingston}, A.~E.,  \& {Dufton}, P.~L. 1986, \mnras, 220, 155

\bibitem[\protect\citeauthoryear{{Keenan} et~al.}{{Keenan}
  et~al.}{2002}]{keenan02}
{Keenan}, F.~P., et~al. 2002, \mnras, 337, 901

\bibitem[\protect\citeauthoryear{{Keenan} et~al.}{{Keenan}
  et~al.}{2009}]{keenan09}
{Keenan}, F.~P., {Crockett}, P.~J., {Aggarwal}, K.~M., {Jess}, D.~B.,  \&
  {Mathioudakis}, M. 2009, \aap, 495, 359

\bibitem[\protect\citeauthoryear{{Landi} et~al.}{{Landi}
  et~al.}{2006}]{chianti5}
{Landi}, E., {Del Zanna}, G., {Young}, P.~R., {Dere}, K.~P., {Mason}, H.~E.,
  \& {Landini}, M. 2006, \apjs, 162, 261

\bibitem[\protect\citeauthoryear{{Liang}, {Badnell}, \& {Zhao}}{{Liang}
  et~al.}{2012}]{liang-b}
{Liang}, G.~Y., {Badnell}, N.~R.,  \& {Zhao}, G. 2012, \aap, 547, A87

\bibitem[\protect\citeauthoryear{{Olluri}, {Gudiksen}, \& {Hansteen}}{{Olluri}
  et~al.}{2013}]{olluri13}
{Olluri}, K., {Gudiksen}, B.~V.,  \& {Hansteen}, V.~H. 2013, \apj, 767, 43

\bibitem[\protect\citeauthoryear{{Peter} et~al.}{{Peter}
  et~al.}{2014}]{peter14}
{Peter}, H., et~al. 2014, Science, 346, 1255726

\bibitem[\protect\citeauthoryear{{Straus}, {Fleck}, \& {Andretta}}{{Straus}
  et~al.}{2015}]{straus15}
{Straus}, T., {Fleck}, B.,  \& {Andretta}, V. 2015, ArXiv e-prints

\bibitem[\protect\citeauthoryear{{Tayal}}{{Tayal}}{1999}]{tayal99}
{Tayal}, S.~S. 1999, Journal of Physics B Atomic Molecular Physics, 32, 5311

\bibitem[\protect\citeauthoryear{{Tayal}}{{Tayal}}{2000}]{tayal00}
{Tayal}, S.~S. 2000, \apj, 530, 1091

\bibitem[\protect\citeauthoryear{{Teriaca}, {Madjarska}, \& {Doyle}}{{Teriaca}
  et~al.}{2001}]{teriaca01}
{Teriaca}, L., {Madjarska}, M.~S.,  \& {Doyle}, J.~G. 2001, \solphys, 200, 91

\bibitem[\protect\citeauthoryear{{Young}, {Feldman}, \& {Lobel}}{{Young}
  et~al.}{2011}]{young11}
{Young}, P.~R., {Feldman}, U.,  \& {Lobel}, A. 2011, \apjs, 196, 23

\end{thebibliography}

\newpage

\appendix

\section{Sanity check}\label{sect.sanity}

\begin{figure}[h]
  \centerline{
    \epsfxsize=3in\epsfbox{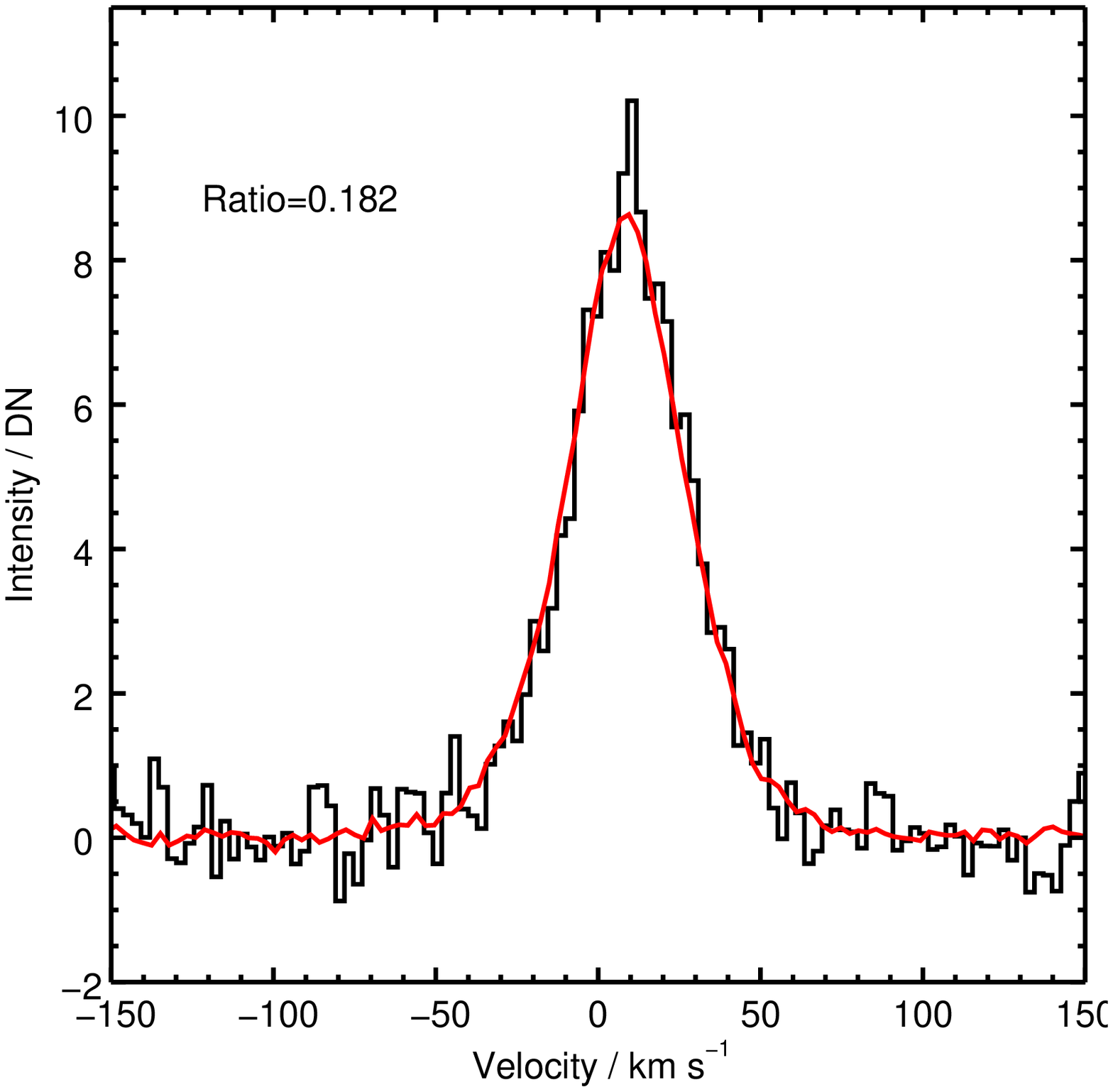}
    \epsfxsize=3in\epsfbox{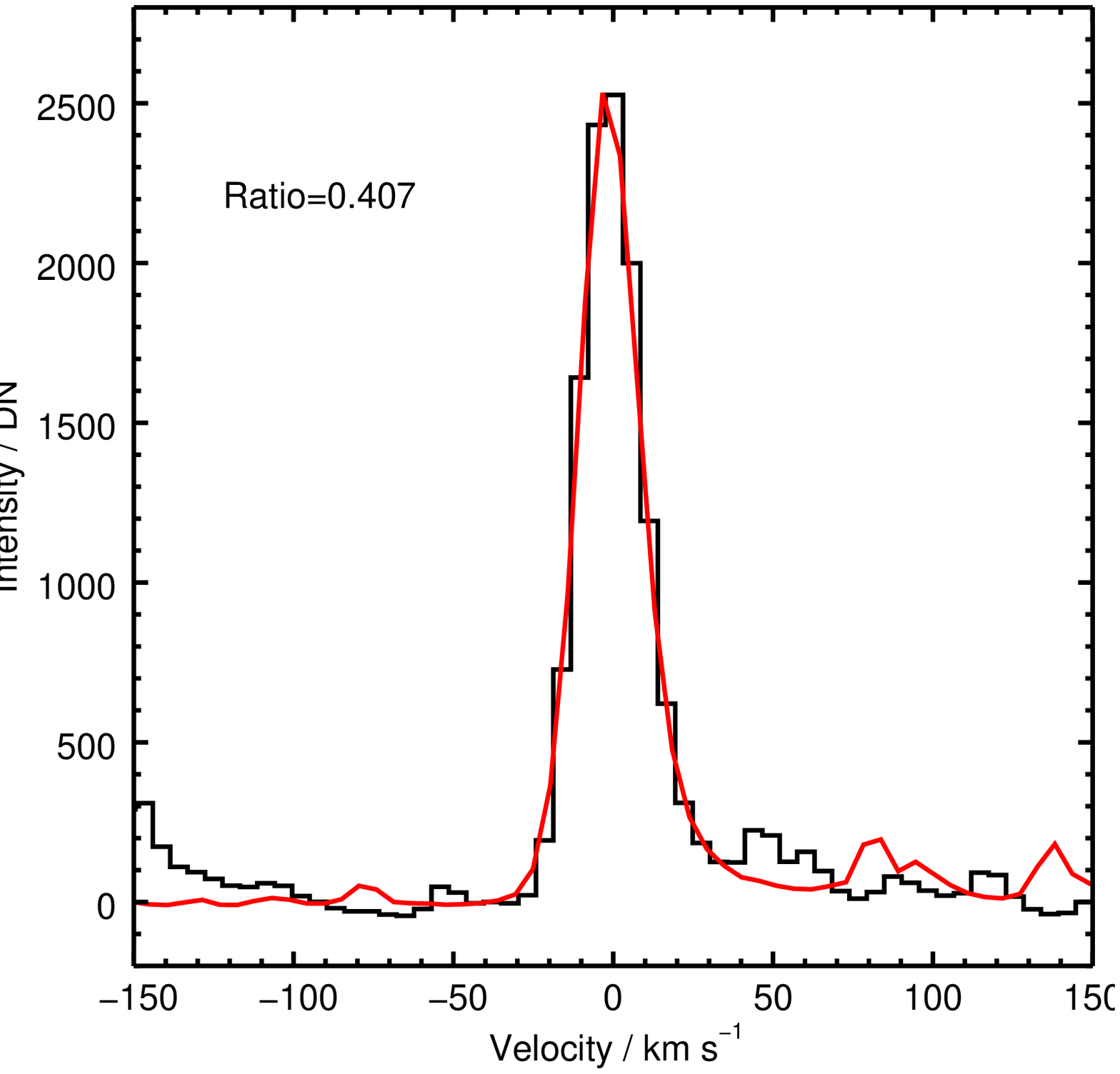}
  }
\caption{A comparison of the \lam1399.8 (black) and \lam1401.2 (red)
  spectral line profiles from spectrum 8 (left) and spectrum 5
  (right). The \lam1401.2 profile has been scaled by the measured
  ratio of the two lines, taken from Table~\ref{tbl.prop}.}
\label{fig.check}
\end{figure}

To demonstrate that the low and high density limits of the \ion{O}{iv}
\lam1399.8/\lam1401.2 ratio are reached in the coronal hole and flare
kernel data-sets, respectively, we show in Figure~\ref{fig.check} the
\lam1399.8 and \lam1401.2 line profiles over-plotted in velocity
space, with the \lam1401.2 profile scaled by the measured line ratio
(Table~\ref{tbl.prop}). 

\end{document}